\documentclass[twocolumn,english,floatfix,superscriptaddress,prl,showpacs]{revtex4}
\usepackage[T1]{fontenc}
\usepackage[latin1]{inputenc}
\usepackage{amsmath}
\usepackage{graphicx}
\usepackage{amssymb}

\usepackage{amsfonts}

\usepackage{amsfonts}

\makeatletter
\makeatother

\usepackage{babel}

\begin{document}

\title{Electronic Griffiths phase of the $d=2$ Mott transition }

\author{E. C. Andrade}

\affiliation{Department of Physics and National High Magnetic Field Laboratory,
Florida State University, Tallahassee, FL 32306}

\affiliation{Instituto de Física Gleb Wataghin, Unicamp, C.P. 6165, Campinas,
SP 13083-970, Brazil}

\author{E. Miranda}

\affiliation{Instituto de Física Gleb Wataghin, Unicamp, C.P. 6165, Campinas,
SP 13083-970, Brazil}

\author{V. Dobrosavljevi\'{c}}

\affiliation{Department of Physics and National High Magnetic Field Laboratory,
Florida State University, Tallahassee, FL 32306}

\date{\today}
\begin{abstract}
We investigate the effects of disorder within the $T=0$ Brinkman-Rice
(BR) scenario for the Mott metal-insulator transition (MIT) in two
dimensions ($2d$). For sufficiently weak disorder the transition
retains the Mott character, as signaled by the vanishing of the local
quasiparticles (QP) weights $Z_{i}$ and strong disorder screening
at criticality. In contrast to the behavior in high dimensions, here
the local \textit{spatial fluctuations} of QP parameters are strongly
enhanced in the critical regime, with a distribution function $P(Z)\sim Z^{\alpha-1}$
and $\alpha\rightarrow0$ at the transition. This behavior indicates
a robust emergence of an electronic Griffiths phase preceding the
MIT, in a fashion surprisingly reminiscent of the {}``Infinite Randomness
Fixed Point'' scenario for disordered quantum magnets.
\end{abstract}


\maketitle
The effects of disorder on quantum criticality \cite{tvojta_jpa}
prove to be much more dramatic than in classical systems. Here, some
critical points can be described by an {}``infinite randomness fixed
point'' (IRFP) \cite{fishertransising2} and the associated quantum
Griffiths phase. Such exotic behavior is well established in insulating
quantum magnets with discrete internal symmetry of the order parameter
\cite{NFL_2005}, but may or may not survive in other models or in
the presence of dissipation due to conduction electrons.

More general insight in the robustness of the IRFP scenario rests
on a recently proposed symmetry classification \cite{tvojta_jpa},
based on the lower critical dimension of droplet excitations. These
ideas have found support in very recent work \cite{hoyos-2007-99},
sparking considerable renewed interest \cite{delmaestro-2008}. Much
of this progress, however, relies on the ability to identify an appropriate
order parameter, describing the corresponding symmetry breaking transitions.

The metal-insulator transition (MIT) represents another important
class of quantum criticality, one that often cannot be reduced to
breaking any static symmetry. Conventional theories of the MIT in
disordered systems \cite{lee_ramakrishnan}, based on the diffusion
mode picture, strongly resemble standard critical phenomena and thus
do not easily allow \cite{NFL_2005} for rare event physics or IRFP
behavior. There currently exists, however, a large body of experimental
work \cite{stewart_nfl_rmp2001}, documenting disorder-induced non-Fermi
liquid behavior due to rare disorder configurations, even in systems
far from any spin or charge ordering.

Theoretically, such {}{}``electronic Griffiths phases\textquotedblright\ (EGP)
\cite{Electronic_Griffths_PRL,Effective_Griffths_PRB,statDMFT} have
recently been proposed for correlated electronic systems with disorder,
based on generalized dynamical mean-field theory (DMFT) approaches
\cite{dmft_rmp96,NFL_2005}. All these works were performed on the
Bethe lattice and identified EGPs only in the vicinity of disorder-driven
MITs, in particular, only for strong enough disorder, in contrast
to quantum magnets where even weak disorder often results in IRFP
behavior. Some key unanswered questions thus remain: (a) What is the
effect of weak to moderate disorder on interaction-driven MITs such
as the Mott transition in finite dimensions? (b) Is the critical behavior
dramatically changed as in the examples of IRFP or may a more conventional
scenario suffice?

\begin{figure}[h]
\bigskip{}

\begin{centering}
\includegraphics[bb=53bp 40bp 714bp 530bp,scale=0.3]{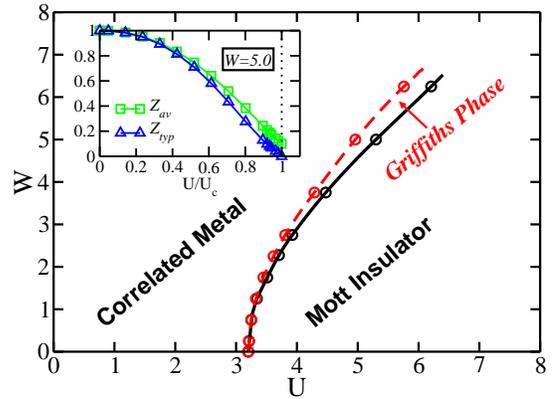} 
\par\end{centering}

\caption{$T=0$ phase diagram of the disordered half-filled Hubbard model in
$d=2$, as a function of the interaction $U$ at weak to moderate
disorder strength $W\lesssim U$. An intermediate Electronic Griffiths
Phase emerges separating the disordered Fermi liquid metal and the
Mott insulator. The inset shows the typical ($Z_{typ}$) and average
($Z_{av}$) values of the local quasiparticle weight $Z_{i}$ as a
function of $U$. The Mott transition is identified by the (linear)
vanishing of $Z_{typ}$. Note that $Z_{av}$ is finite at $U_{c}$,
indicating that a fraction of the sites remains nearly empty or doubly
occupied. }

\label{fig:1}
\end{figure}

In this Letter, we investigate the effects of weak and moderate disorder
on the Mott MIT at half filling \cite{mott_mit} in two dimensions.
As the simplest description of the effects of disorder on the Mott
transition, we work within a Brinkman-Rice (BR) scenario \cite{Brinkman_Rice},
where a Gutzwiller variational approximation is applied to a disordered
two dimensional Hubbard model. Our results demonstrate that: (i) for
sufficiently weak disorder the transition retains the second order
Mott character, where electrons gradually turn into localized magnetic
moments; (ii) disorder-induced spatial inhomogeneities give rise to
an \textit{intermediate} EGP that displays IRFP character at criticality,
even when the transition is approached by increasing the interaction
at weak disorder; (iii) the renormalized disorder seen by quasiparticles
is strongly screened only at low energies, resulting in pronounced
energy-resolved inhomogeneity of local spectral functions.

\emph{Model.}---We focus on the paramagnetic disordered Hubbard model
with nearest-neighbor hopping and with site energies $\varepsilon_{i}$
uniformly distributed in the interval $\left[-W/2\mbox{,}\, W/2\right]$
\cite{statDMFT,Screening_2003}. We approach the Mott transition by
increasing the on-site Hubbard interaction $U$ at half filling (chemical
potential $\mu=U/2$), on an $L$x$L$ square lattice with periodic
boundary conditions. All energies will be expressed in units of the
clean Fermi energy (half-bandwidth) $E_{F}=4t$, where $t$ is the
hopping amplitude.

Within our disordered BR approach, we self-consistently calculate
the local single-particle self-energies $\Sigma_{i}(\omega_{n})$
\cite{statDMFT,Screening_2003}, which assume a site-dependent form
\begin{equation}
\Sigma_{i}\left(\omega_{n}\right)=\left(1-Z_{i}^{-1}\right)\omega_{n}+v_{i}-\varepsilon_{i}+\mu.\label{eq:self_energy_SB4}\end{equation}
 The renormalized site energies $v_{i}=v_{i}\left(e_{i},d_{i}\right)$
and the local quasiparticle (QP) weights $Z_{i}=Z_{i}\left(e_{i},d_{i}\right)$
are variationally calculated through the saddle-point solution of
the corresponding Kotliar-Ruckenstein (KR) slave boson functional
\cite{KRSB4}\begin{align}
F & =-2T\sum_{\omega_{n}}Tr\ln\left[-i\omega_{n}\mathbf{1}+\mathbf{Z}\mathbf{v}+\mathbf{Z}^{1/2}\mathbf{H}_{0}\mathbf{Z}^{1/2}\right]\nonumber \\
 & +\sum_{i}\left[Ud_{i}^{2}-\left(1-e_{i}^{2}+d_{i}^{2}\right)\left(Z_{i}v_{i}-\varepsilon_{i}+\mu\right)\right].\label{eq:Free_energy_SB4}\end{align}
 Here, $e_{i}$ and $d_{i}$ are the KR slave boson amplitudes \cite{KRSB4},
$T$ is the temperature, and $\omega_{n}$ are the Matsubara frequencies.
The operators $\mathbf{Z}$ and $\mathbf{v}$ are site-diagonal matrices
$[\mathbf{Z}]_{ij}=Z_{i}\delta_{ij}$; $[\mathbf{v}]_{ij}=v_{i}\delta_{ij}$,
and $\mathbf{H}_{0}$ is the clean and non-interacting lattice Hamiltonian.

\begin{figure}[h]
\bigskip{}

\begin{centering}
\includegraphics[bb=53bp 40bp 714bp 530bp,scale=0.3]{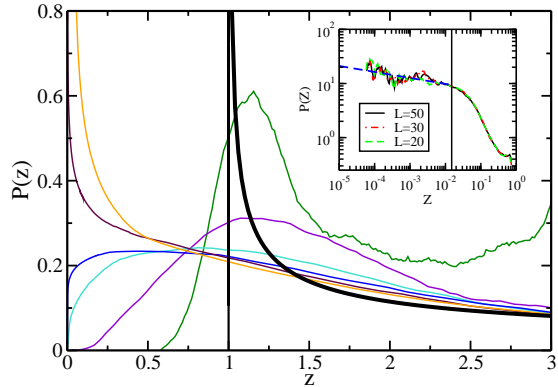} 
\par\end{centering}

\caption{Distribution of $z_{i}=Z_{i}/Z_{0}$ (see text) as the transition
is approached by increasing $U$ (thin lines correspond to $U/U_{c}(W)=0.6,0.8,0.9,0.92,0.94,0.97$).
For reference, the thick solid line shows the DMFT {}``fixed point''
distribution which remains bounded from below. Our $d=2$ results
show that, due to rare events, a low-$z$ tail $P\left(z\right)\sim z^{\alpha-1}$
emerges for $z\lesssim1$ as the transition is approached. In the
critical region the distribution assumes a singular form ($\alpha<1$),
indicating the onset of an Electronic Griffiths Phase. Results are
shown for $W=5.0$ and $L=20$. The inset illustrates how for such
large lattices our results for $P\left(Z\right)$ are essentially
independent of the system size (shown for $U/U_{c}(W)=0.94$ and $\alpha=0.77\pm0.06,0.81\pm0.05,0.82\pm0.05$
corresponding to $L=20,30,50$, respectively). }

\label{fig:2}
\end{figure}

This approach is mathematically equivalent to a generalization of
the dynamical mean field theory (DMFT) \cite{dmft_rmp96} to finite
dimensions, the \textit{``statistical DMFT}''\cite{statDMFT} implemented
using a slave boson impurity solver, which provides an elegant and
efficient computational approach, allowing us, for example, to calculate
$Z_{i}$ values spanning eight orders of magnitude. We considered
several lattice sizes ranging up to $L=50$, and for every $\left(U,W\right)$
pair we typically generated around forty realizations of disorder.
We carefully verified that for such large lattices, all our results
are robust and essentially independent of the system size (see, e.g.,
the inset of Fig. \ref{fig:2}).

\emph{Phase diagram and Griffiths phase.}---To characterize the $T=0$
disordered Mott transition in $d=2$, we follow the evolution of the
local QP weights $Z_{i}$, as the interaction $U$ is increased at
fixed disorder $W$. For weak to moderate disorder, we find behavior
partly reminiscent of that previously established for high dimensions
({}{}``DMFT limit\textquotedblright) \cite{effects_dis_prb_2005}.
The approach to the critical point at $U=U_{c}(W)$ is identified
by the vanishing of the typical QP weight $Z_{typ}=\exp\{\left\langle \ln Z_{i}\right\rangle \}$\cite{statDMFT},
indicating the Mott transmutation of a finite (large) fraction of
electrons into local magnetic moments. Because random site energies
tend to push the local occupation away from half filling $U_{c}(W)$
increases with disorder (Fig. \ref{fig:1}).

The role of fluctuation effects, however, is best seen by contrasting
our $d=2$ results to those found in the DMFT limit \cite{Screening_2003}.
There each site has many neighbors, and thus {}``sees'' the same
(self-averaged) environment ({}``cavity''), so $Z_{i}^{DMFT}=Z^{DMFT}\left(\varepsilon_{i}\right)$
depends only on the local site energy $\varepsilon_{i}$. Its minimum
value corresponds to the sites closest to half-filling $Z_{\min}^{DMFT}\equiv Z_{o}=Z^{DMFT}\left(\varepsilon_{i}=0\right)$.
In the critical region all $Z_{i}^{DMFT}\sim\left\langle Z\right\rangle \sim U_{c}^{DMFT}(W)-U$,
but the scaled local QP weights $z_{i}^{DMFT}=Z_{i}^{DMFT}/Z_{o}$
approach finite values at the transition, with $z_{\min}^{DMFT}=1$.
The corresponding scaled distribution $P\left(z_{i}^{DMFT}\right)$
approaches a fixed-point form close to $U_{c}^{DMFT}(W)$ (shown by
the thick solid line in Fig. \ref{fig:2}).

In low dimensions, site-to-site cavity fluctuations give rise to a
low-$Z$ tail emerging below the DMFT minimum value $Z_{o}$ (Fig.
\ref{fig:2}). To bring this out, we present our $d=2$ results in
precisely the same fashion as in the DMFT limit, i.e. scaling each
$Z_{i}$ with $Z_{o}$. Away from the transition the distribution
resembles the DMFT form, but in the critical region the low-$z$ tail
assumes a power-law form \begin{equation}
P\left(z\right)\sim z^{\alpha-1}.\label{eq:P(z)}\end{equation}

Physically, the emergence of a broad distribution of local QP weights
$Z_{i}$ indicates the presence of rare disorder configurations characterized
by anomalously low local energy scales $\Delta_{i}\approx Z_{i}E_{F}$.
Since the approach to the Mott insulator corresponds to $Z\rightarrow0$,
such regions with $Z_{i}\ll Z_{typ\text{ }}$should be recognized
as {}``almost localized'' Mott droplets. Within our BR picture,
each local region provides \cite{statDMFT} a contribution $\chi_{i}\sim\gamma_{i}\sim Z_{i}^{-1}$
to the spin susceptibility or the Sommerfeld coefficient, respectively.
The local regions with the smallest $Z_{i}$ thus dominate the thermodynamic
response and produce non-Fermi liquid metallic behavior \cite{Electronic_Griffths_PRL,Effective_Griffths_PRB}
whenever $\alpha<1$.

\begin{figure}[h]
\bigskip{}

\begin{centering}
\includegraphics[bb=53bp 40bp 714bp 530bp,scale=0.3]{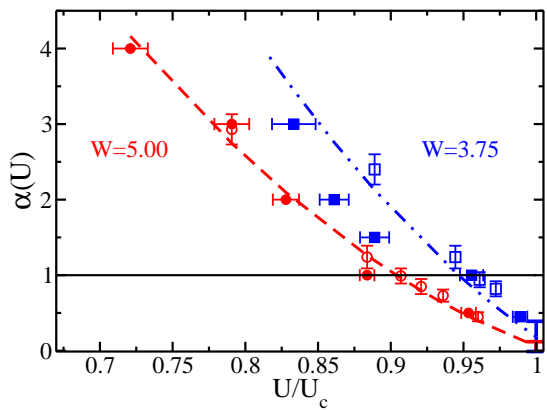} 
\par\end{centering}

\caption{Exponent $\alpha\left(U\right)$ for two different values of disorder.
To calculate $\alpha$ we use two methods (open and closed symbols,
see text). Result are shown for $L=20$. Within the estimated error
bars, we find an extrapolated value consistent with $\alpha=0$ at
the critical point $U=U_{c}(W)$.}

\label{fig:3}
\end{figure}

\emph{IRFP-like behavior.}---To carefully calculate the exponent $\alpha(U)$
as the transition is approached, we use two distinct methods. The
first relies on the {}``estimator'' \cite{power_law_estimator}
$\alpha=\left\langle \ln\left[Z_{max}/Z_{i}\right]\right\rangle _{Z_{i}\leq Z_{max}}^{-1}$,
where $Z_{max}\sim Z_{typ}\sim Z_{o}$ (see Fig. \ref{fig:2}) is
an appropriate upper-bound on the power-law behavior. The second approach
consists in calculating $\left\langle Z^{-\mu}\right\rangle _{Z_{i}\leq Z_{max}}^{-1}$
for given $\mu$, as a function of $U$. This quantity is expected
to vanish at $U=U_{\mu}$, satisfying $\alpha(U_{\mu})=\mu$. Both
methods give consistent results (open and closed symbols, respectively,
in Fig. \ref{fig:3}), which agree within the estimated error bars.

The exponent $\alpha$ is found to decrease smoothly as the transition
is approached, until the distribution assumes a singular form ($\alpha<1$),
indicating the emergence of an Electronic Griffiths Phase (EGP) \cite{Electronic_Griffths_PRL,Effective_Griffths_PRB}.
Its estimated onset ($\alpha=1$) is generally found to strictly precede
the MIT (dashed line in Fig. \ref{fig:1}), indicating that disorder
fluctuations qualitatively modify the critical behavior even for weak
to moderate disorder. Remarkably, we find that, within our numerical
accuracy, $\alpha\rightarrow0$ precisely along the critical line
$U=U_{c}(W)$! This establishes a phenomenology which closely parallels
the behavior of magnetic Griffiths phases with IRFP behavior \cite{tvojta_jpa,NFL_2005}.

\begin{figure}[h]
\begin{raggedright}
\includegraphics[scale=0.5]{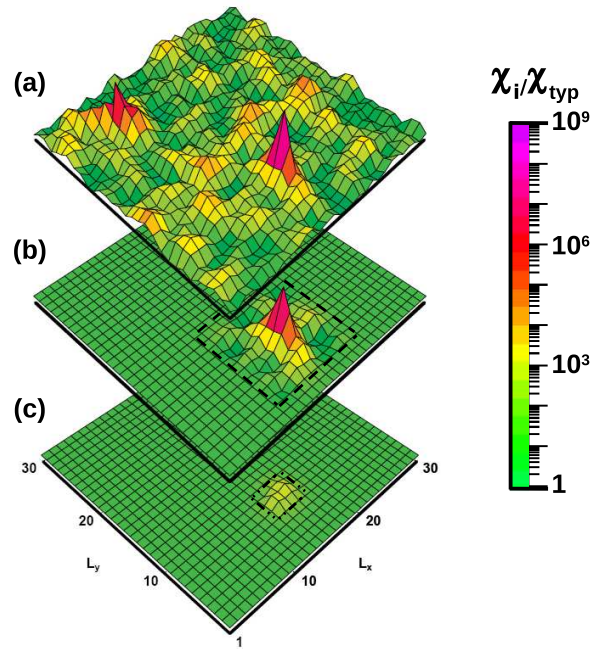} 
\par\end{raggedright}

\caption{(a) Spatial distribution of the (normalized) local spin susceptibilities
$\chi_{i}\sim Z_{i}^{-1}$, illustrating a typical disorder realization
containing a {}``rare event'' (RE) with $\chi_{i}\gg\chi_{typ}$;
(b) Disorder fluctuations are eliminated outside a box of size $l=9$,
without appreciably affecting the RE; (c) When the box is further
reduced (here $l=3$) the RE is rapidly (exponentially) suppressed,
establishing the non-local nature of the rare event, in strong support
of the IRFP picture. Results shown for $L=30$, $W=5.0$ and $U/U_{c}=0.96$.}

\label{fig:4}
\end{figure}

\emph{Structure of the rare events}.---To explore the nature of the
rare events (REs) dominating the EGP, i.e. the regions with $Z_{i}\ll Z_{typ}$,
we examined a number of disorder realizations and selected those few
samples containing the smallest $Z_{i}$. A typical example is shown
in Fig. \ref{fig:4}, where the RE is seen as a very sharp peak of
the local spin susceptibility $\chi_{i}\sim Z_{i}^{-1}$ (note the
logarithmic scale). The corresponding RE site is then placed in the
middle of a box of side $l$ (dashed line in Fig. \ref{fig:4}). To
examine the spatial correlations, we preserve the same disorder realization
within this box, while the outside is replaced by an appropriate DMFT
effective medium. We then recalculate the QP parameters $Z_{i}$ as
the box size is reduced from $l=L$ (original model), down to $l=1$
(DMFT limit where all spatial correlations are suppressed). We find
that the RE is essentially unmodified until the box size reaches $l\sim l_{RE}$
($\approx9$ for the example in Fig. 4), and then is rapidly (exponentially)
suppressed for $l<l_{RE}$. We also find that the variance of the
disorder strength within the box of size $l_{RE}$ is appreciably
weaker than on the average, establishing that the REs dominating the
Griffiths phase stem out from rare disorder configurations, precisely
as expected within the IRFP scenario.

\emph{Critical behavior of the spatial inhomogeneity}.---As $\alpha\rightarrow0$
in the critical region, the $P(Z_{i})$ distribution becomes {}``infinitely
broad'', since $\alpha^{-1}$ measures \cite{NFL_2005} the variance
of $\ln Z_{i}$. The thermodynamic response becomes increasingly \textit{inhomogeneous}
as the transition is approached; such behavior is typically seen in
NMR experiments on materials displaying disorder-driven NFL behavior
\cite{NFL_2005}.

\begin{figure}[h]
\begin{centering}
\includegraphics[scale=0.4]{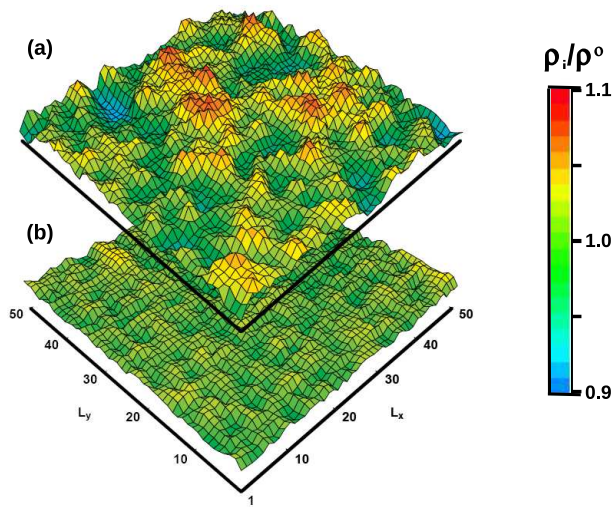} 
\par\end{centering}

\caption{Spatial distribution of the LDOS normalized by its clean value, for
one disorder realization, shown: (a) away from the Fermi energy ($\omega=0.10$);
(b) at the Fermi energy ($\omega=0$). Due to static disorder screening,
the distribution becomes homogeneous close to the Fermi level, but
displays pronounced spatial structures at higher energies. Results
shown for $L=50$, $U/U_{c}=0.96$ and $W=0.75$. }

\label{fig:5}
\end{figure}

But what to expect from STM experiments directly measuring the local
electronic spectra? Within our BR approach, the local density of states
(LDOS) in question $\rho_{i}(\omega)=\frac{1}{\pi}\operatorname{Im}G_{ii}(\omega-i0^{+})$
depends not only on the local QP weights $Z_{i}$, but also on the
renormalized site energies $v_{i}$ through\begin{equation}
G_{ii}(\omega)=\left[\left(\mathbf{Z}^{-1}\omega-\mathbf{v}-\mathbf{H}_{0}\right)^{^{-1}}\right]_{ii}.\label{eq:G_lattice}\end{equation}
 The quasiparticles thus {}``see'' a frequency-dependent effective
disorder potential \begin{equation}
\varepsilon_{i}^{eff}(\omega)=v_{i}-\omega/Z_{i}.\label{eq:effec_dis_pot}\end{equation}

Our explicit calculations find that the renormalized site energies
$v_{i}$ become strongly screened near the transition, giving rise
to a very small (but finite) renormalized disorder strength $W_{eff}=\sqrt{\left\langle v_{i}^{2}\right\rangle }\ll1$
at criticality (e.g. for $W=5,W_{eff}\approx0.05$). Near the Fermi
energy ($\omega=0$), we predict the LDOS spectra to appear increasingly
homogeneous in the critical region. At higher energies, however, the
very broad distribution of local QP weights (essentially local QP
bandwidths) creates a very strong effective disorder seen by the quasiparticles,
and we expect the system to appear more and more \textit{inhomogeneous}
as criticality is approached. This result is illustrated by explicit
computation of the DOS profile (Fig. \ref{fig:5}), which is surprisingly
reminiscent of recent spectroscopic images on doped cuprates \cite{seamus_davis_sci2005}.
Our theory, which does not include any physics associated with superconducting
pairing, strongly suggests that such energy-resolved inhomogeneity
is a robust and general feature of disordered Mott systems.

\emph{Conclusions.}--We presented the first detailed model calculation
investigating the effects of moderate disorder on the Mott metal insulator
in two dimensions. Our findings indicate that rare disorder fluctuations
may dominate quantum criticality even in absence of magnetic ordering
- an idea that begs experimental tests on a broad class of materials.
The Brinkman-Rice scenario we considered, which focuses on local (Kondo-like)
effects of strong correlation (while neglecting inter-site magnetic
correlations), may be relevant only for systems with sufficiently
strong magnetic frustration, such as $^{3}\mbox{He}$ monolayers adsorbed
on graphite \cite{saunders_3He}. Such variational approach should
be generalized for systems, such as copper oxides, where the inter-site
super-exchange is strong, but this fascinating research direction
remains a challenge for future work.

This work was supported by FAPESP through grant 04/12098-6 (ECA),
CAPES through grant 1455/07-9 (ECA), CNPq through grant 305227/2007-6
(EM), and by NSF through grant DMR-0542026 (VD).


\end{document}